\documentclass{article}
\usepackage[left=1.85cm, right=2.25cm,top=2.5cm,bottom=3cm]{geometry}
\usepackage{graphicx}
\usepackage[greek,english]{babel}
\usepackage{natbib}
\usepackage[resetlabels]{multibib}
\newcites{further}{Further Reading}
\usepackage{amsmath}
\usepackage{placeins}
\usepackage{xcolor}
\usepackage{color}
\usepackage{colortbl}
\usepackage{multirow}
\usepackage{soul}
\graphicspath{{Figures/}}
\title{Market Regime Detection via Realized Covariances: A Comparison between Unsupervised Learning and Nonlinear Models}
\author{Andrea Bucci$^1$ \and Vito Ciciretti$^2$}
\date{
$^1$Department of Economics, Universit\`a degli Studi "G. d'Annunzio" Chieti-Pescara\\
$^2$Independent Researcher}

\sethlcolor{yellow}

\makeatletter
\def\SOUL@hlpreamble{%
    \setul{}{3.5ex}
    \let\SOUL@stcolor\SOUL@hlcolor
    \dimen@\SOUL@ulthickness
    \dimen@i=-.75ex 
    \advance\dimen@i-.5\dimen@
    \edef\SOUL@uldepth{\the\dimen@i}%
    \let\SOUL@ulcolor\SOUL@stcolor
    \SOUL@ulpreamble
}
\makeatother
\usepackage{etoolbox}

\usepackage[colorlinks]{hyperref}

\makeatletter
\newcommand{\sectionbiblio}{%
  \patchcmd{\std@thebibliography}{\chapter*}{\section*}{}{}
}
\makeatother

\makeatletter
\patchcmd{\NAT@test}{\else \NAT@nm}{\else \NAT@nmfmt{\NAT@nm}}{}{}

\DeclareRobustCommand\citepos
  {\begingroup
   \let\NAT@nmfmt\NAT@posfmt
   \NAT@swafalse\let\NAT@ctype\z@\NAT@partrue
   \@ifstar{\NAT@fulltrue\NAT@citetp}{\NAT@fullfalse\NAT@citetp}}

\let\NAT@orig@nmfmt\NAT@nmfmt
\def\NAT@posfmt#1{\NAT@orig@nmfmt{#1's}}

\makeatother

\begin{document}
\maketitle
\begin{abstract}
There is broad empirical evidence of regime switching in financial markets. The transition between different market regimes is mirrored in correlation matrices, whose time-varying coefficients usually jump higher in highly volatile regimes, leading to the failure of common diversification methods. In this article, we aim to identify market regimes from covariance matrices and detect transitions towards highly volatile regimes, hence improving tail-risk hedging. Starting from the time series of fractionally differentiated sentiment-like future values, two models are applied on monthly realized covariance matrices to detect market regimes. Specifically, the regime detection is implemented via vector logistic smooth transition autoregressive model (VLSTAR) and through an unsupervised learning methodology, the agglomerative hierarchical clustering. Since market regime switches are unobservable processes that describe the latent change of market behaviour, the ability of correctly detecting market regimes is validated in two ways: firstly, randomly generated data are used to assess a correct classification when regimes are known; secondly, a na\"{i}ve trading strategy filtered with the detected regime switches is used to understand whether an improvement is showed when accounting for regime switches. The results point to the VLSTAR as the best performing model for labelling market regimes.

\end{abstract}

\newpage
\section{Introduction}

Financial markets are often seen changing behaviour abruptly. Trying to statistically describe financial markets’ movements often translates in changing the level of some parameters such as the mean of the returns, their volatility, the correlation between assets, and so on. For instance, codependence measures such as linear correlations dramatically changed at the start of the 2008 Great Financial crisis or at the inception of the COVID-19 crisis which chiefly led to a quick risk-off phase in 2020. Market regime switches are unobservable processes that describe the latent change of market behaviour and may have different persistency. While regime switches may be present both at higher and lower frequencies, in this paper we focus more on short term regime switches rather than long-term ones. The original idea of regime switches was laid out by \citet{hamilton1988} who linked business cycle regimes to cycles of economic activity. In his paper, the author describes the term structure of three-month treasury bills and ten-year bonds with discrete shifts in the parameters governing the behaviour of exogenous variables. In addition, the author models the US real GDP with an autoregressive specification whose parameters switch on a discrete Markov process, at the same time presenting an algorithm for deriving the Markov probabilistic structure. 


The idea of modelling regime switches is appealing as it allows to account for several statistical properties of financial markets such as fat tails, volatility clustering, skewed distributions, time-varying correlations and non-linear relationships. Nevertheless, detecting and predicting structural breaks can play a key role not only in better describing the statistical properties of financial markets but also can contribute to asset pricing, risk management and portfolio construction. For instance, since in highly volatile regimes both the absolute average correlation and the correlation between and within industries usually jump higher, classical diversification often fails when it is most needed. Also, classical asset allocation methods based on the na\"{i}ve inverse of the covariance matrix will likely fail due to the higher condition number proper of highly volatile market regimes. \citet{ang2011} describe the basic structure of regime-switching models as well as the statistical advantage of incorporating regime switches whose non-linearity allows to describe several empirical facts of financial variables such as fat tails, ARCH effects, skewness and time-varying correlations.

\citet{guidolin2011} reviews the literature of Markov regime-switching models in empirical finance to validate multiple priors related to these models. Most importantly, the author first questions the usefulness of Markov switching models for statistical or economic reasons, not finding evidence in favour of any of the two. Finally, the author shows that Markov switching models are most successfully applied at monthly frequency and may help to improve the long-term predictability of fundamental idiosyncratic quantities. In a similar framework, \citet{case2014} apply Markov Switching models to model REIT returns. 

The strand of literature focusing on regime detection includes both models based on stochastic process drifts and on volatility changes. For example, \citet{yuan2016} use a Hidden Markov model to govern regime-switches in the parameters of a Geometric Brownian motion. In the same context, \citet{chevallier2014} detect regime-switches in several international stock indexes by means of a Lévy process specified with a drift, a scaled Brownian motion and an independent pure jump process. Regimes are proxied by the Lévy jump component and are detected by means of a Normal Inverse Gaussian distribution fitted to each regime. 
\citet{Liu2020} apply GARCH and HAR-RV Markov switching models to detect regime breaks in the volatility of oil prices concluding that the implied transition rate pins-down the long-term structure of volatility. \citet{Tsang2018} applies a data-driven approach based on directional change to develop an indicator capable to detect Brexit related regime changes. \citet{Chun2013} apply a regime detection technique based on a sequential t-test to credit spreads and find two distinct regimes, a long-lived level regime linked to the Federal Reserve policy and a short-lived volatility regime detected during major financial crises. Regime switches have also been applied by the asset allocation literature. \citet{kim2012} applies a three regimes Hidden Markov model for a dynamic asset allocation problem yielding an improved out-of-sample performance compared to a basic cap-weighted allocation. More recently, \citet{fons2021} build a dynamic asset allocation system using Feature Saliency Hidden Markov models to detect regime switches.

In this manuscript, we aim to identify two market regimes – a calm and a highly-volatile one – starting from monthly realized covariance matrices. Once calculated the realized covariance matrices from the time series of fractionally differentiated sentiment-like future values, we model them through a vector logistic smooth transition autoregressive model (VLSTAR) and an unsupervised learning methodology, the agglomerative hierarchical clustering, to identify two latent market regimes. Both models are suitable to regime detection as while the hierarchical clustering bases the regime detection on patterns dependent on the distance between variables, the VLSTAR is based on a cumulative logistic function which is capable of translating the movements of a transition variable into a probabilistic framework. To further understand regime changes through a multivariate model, we also apply a Threshold vector autoregressive model (TVAR). To the best of our knowledge, this is the first study that implements a vector logistic smooth transition autoregressive models (VLSTAR) and hierarchical clustering to the problem of market regime detection. VLSTAR models allow to specify an exogenous transition variable that governs the regime switches. This is often empirically observed in financial markets, especially with less systematic financial variables which are still co-dependent on more systematic sources of risks. On the other hand, hierarchical clustering is an unsupervised learning technique that tries to label the latent regimes by means of changes in distances inter and intra-clusters. We choose to detect regime switches at monthly frequency to allow the underlying mean-reverting data generating processes to have deviations that are not regarded as structural breaks. 

Since market regimes’ switches are a latent unobservable process, to assess the validity of our approach we implemented a validation framework based on two pillars. First, we apply the three models to a randomly generated time series with controlled variance. In fact, by keeping the mean stable and by modifying the variance of a random data generation process, we are capable of exactly identifying which periods belong to a highly volatile regime and which to a calm one. Second, we compare the performance of a na\"{i}ve momentum investment strategy which empirically performs well during calm regimes with its regime-filtered version which does not allow trades when a transition to a highly volatile regime is spotted. As such, we expect the investment strategy not only avoiding long equity positions during market downturns, but also to diminish the downside volatility and to reduce the exposure to liquidity crises.


The rest of the paper is organized as follows. Section 2 offers an in-depth description of the VLSTAR and Hierarchical clustering models. Section 3 describes the data and the results of the training process. Section 4 deals with the validation framework. Section 5 concludes and addresses potential future work.

\section{Methodology}

\subsection{VLSTAR}

The first model we employ to detect regime switches is the Vector Smooth Transition Autoregressive (VLSTAR) model with lagged exogenous variables. This class of models explicitly assumes that the regime switch is determined by an observable transition variable. The use of this kind of model on covariances is coherent with the empirical finding that volatility and covariances are time-varying and particularly linked to news outflow in different market regimes. In this paper we employ the multivariate specification firstly appeared in \citet{ty14}.

The relationship between  $n$  assets elements of  $y_t$  is given by:

\begin{align*}
y_t&=\mu _0+\sum _{j=1}^p\varphi _{0,j}y_{t-j}+A_0x_t+G_t\left(s_t;\gamma ,c\right)\left[\mu _1+\sum _{j=1}^p\varphi _{1,j}y_{t-j}+A_1x_t\right]+\epsilon_t\\
&=\mu _0+G_t\left(s_t;\gamma ,c\right)\mu _1+\sum _{j=1}^p\left[\varphi _{0,j}+G_t\left(s_t;\gamma ,c\right)\varphi _{1,j}\right]y_{t-j}+[A_0+G_t\left(s_t;\gamma ,c\right)A_1]x_t+\epsilon_t.
\end{align*}
Where  $\mu _0$  and  $\mu _1$  are the  $n+1$  vectors of intercepts,  $\varphi _{0,j}$  and  $\varphi _{1,j}$  are  $n \times  n$  square matrices of parameters for lags  $j=1,2,{\ldots},p$,  $A_0$  and  $A_1$  are  $n \times k$  matrices of parameters,  $x_t$  is the  $k \times 1$ vector of exogenous variables and  $\epsilon _t$ is the vector of innovations. Most importantly,  $G_t\left(s_t;\gamma, c\right)$  is an  $n \times n$  diagonal matrix of transition function at time  $t$, such that: 

\begin{equation*}
G_t\left(s_t;\gamma ,c\right)=\mathit{diag}\{G_{1,t}\left(s_{1,t};\gamma _1,c_1\right),G_{2,t}\left(s_{2,t};\gamma _2,c_2\right),{\ldots},G_{n,t}\left(s_{n,t};\gamma _n,c_n\right)\}.
\end{equation*}

Every scalar transition function  $G_{i,t}\left(s_{i,t};\gamma _i,c_i\right)$  with  $i=1,2,{\ldots},n$  is a continuous function of the transition variables  $s_{i,t}$  with scale parameter  $\gamma _i$  and threshold  $c_i$.

The model can be extended to include  $m-1$  regime switches, hence becoming: 

 \begin{align*}
    y_t&=\mu _0+\sum _{j=1}^p\varphi _{0,j}y_{t-j}+A_0x_t+G_t\left(s_t;\gamma ,c\right)\left[\mu _1+\sum _{j=1}^p\varphi _{1,j}y_{t-j}+A_1x_t\right]+{\dots}+\\
    &+G_t^{m-1}\left(s_t;\gamma ,c\right)\left[\mu _1+\sum _{j=1}^p\varphi _{1,j}y_{t-j}+A_1x_t\right]+\epsilon_t.
   \end{align*}

The model is estimated via non-least squares. The key component of a vector smooth transition model is the transition function  $G_t$, which is bounded between zero and one. When it assumes the value of zero, then the baseline model becomes a linear VAR. Hence,  $G_t$  may be interpreted as a filtering rule that locates the model between two extreme regimes.

The diagonal element of  $G_{i,t}^r$ are specified as a logistic cumulative density function with a unique transition variable and common $\gamma$ and $c$, such that:

\begin{equation*}
G_{i,t}^r\left(s_{t}^r;\gamma^r,c^r\right)=\left[1+\exp \left\{-\gamma^r(s_{t}^r-c^r)\right\}\right]^{-1}
\end{equation*}

for  $i=1,2,{\dots},n$  and $r =1,2,{\dots},(m-1)$. The  $\gamma $  parameter controls the slope, hence determining the shape of the logistic function and the smoothness of the transition. When   $\gamma = 0$ , then the equation becomes linear, while for  $\gamma \rightarrow {\infty}$  the transition becomes abrupt. The  $c$  parameter, instead, is a location parameter determining the mid-point of the transition. In this case, we are supposing a common logistic function, with only two parameters to estimate for each regime (i.e. $\gamma$ and $c$) for all the dependent variables.

To understand which variable use as single transition variable, we first test linearity for all the candidates driving factors. Thus, we test linearity for each of the potential variables and selected the one exhibiting the lowest 
\emph{p}-value. The linearity test we implement is the one discussed in \citet{luukkonen1988} and \citet{granger1993}, and extended to the multivariate framework by \citet{teraasvirta2010}.

In the specification of the market regime strategy, we also define that the desirable number of regimes to be detected, $m$, as equal to two. This decision is mainly motivated by the empirical evidence that volatility moves between highly volatile periods and low volatility periods, see also \citet{davies1987}, \citet{hansen1992}, \citet{cho2007}. As such, we choose to detect two market-regimes: a calm regime and a highly volatile one. To this end, we use the logistic function, $G_t$. Specifically, we assume a given regime for the $t$-observation when the diagonal elements of $G_t$ are equal or lower than $0.5$ and the other regime is detected for values greater than $0.5$.

\subsection{Hierarchical Clustering}

Clustering analysis is an unsupervised learning technique aims at grouping a collection of objects into subsets such that the objects within each cluster are more closely related than to objects assigned to different clusters. The application to time series helps to identify structural similarities in different moments and space by performing coarse-graining so that stable relationships appear between clusters and not between individual time series. 

In this work, we employ hierarchical clustering - first introduced in \citet{Johnson1967} - as it allows us to find groupings with well-defined hierarchies within relationships. The latter characteristic may be particularly useful in financial markets, as investment decisions are often based on hierarchical cause-effect relationships, such as changes in monetary policies affecting the discount function and in turn fixed income and equity prices. There exist two paradigms for hierarchical clustering. The agglomerative one starts with assuming that each object belongs to a different cluster, to then merge at each of the  $N-1$  steps the two least dissimilar clusters in a single cluster by minimizing the inter-cluster dissimilarity. The divisive hierarchical clustering, instead, takes the opposite stance by assuming that all objects belong to the same cluster and to then split them into different ones by maximizing the between-clusters dissimilarity.

We use the agglomerative paradigm by means of the AGNES algorithm of \citet{Kaufman1990} due to the elevated computational complexity associated with the divisive algorithm as it requires  $2^{n-1}-1$  ways of splitting a set of  $N$  objects into two subsets. We consider the features  $x_{t,j}$  for  $t=1,{\dots},T$  and  $j=1,{\dots},N$  extracted from the realized correlation matrices as the base for measuring intra-cluster dissimilarities. Since a degree of (dis-)similarity is the central notion in clustering, we start by defining the pairwise dissimilarities as: 

\begin{equation*}
D\left(x_j,x_{j^{\prime}}\right)=\sum _{t=1}^Td_t(x_{t,j},x_{t,j^{\prime}})
\end{equation*}

where  $d$  is a distance function between each pair of observations. Each distance function implements a specific concept of similarity and must be a metric. In this case, $d$ is specified as the Manhattan distance in which the distance between two points is the sum of the absolute differences of their Cartesian coordinates.

Once identified the first clusters through the Manhattan distance, groups are further merged relying on a second measure of dissimilarity between-clusters. Let  $G$  and  $H$  represent two of such groups. The dissimilarity between the two groups  $d(G,H)$  is computed from the set of pairwise dissimilarities  $d_{j,j^{\prime}}$ where $j{\in}G$  and  $j^{\prime}{\in}H$  by means of a linkage function. Here, $d_{j,j^{\prime}}$ is defined as a Ward linkage function which analyses the variance of the clusters via the sum of square distances from the centroids, instead of measuring the distance directly.

\section{Data and in-sample results}

We use hourly prices of 9 futures contracts for the period January 2010 - October 2020, sourced from Bloomberg, for a total of $T^*=33,200$ hourly observations, to build $T=560$ monthly realized covariance matrices. Table~\ref{tab:1} reports the underlying names and the CME future codes. These assets have been chosen due to their empirical capability to describe the overall market sentiment.

\begin{table}
  \centering
  \caption{List of the securities composing the dataset and their CME future codes}\label{tab:1}
  \begin{tabular}{lc}

    \bfseries Name & \bfseries CME Code\\
      	\hline
S\&P500 & SP \\
VIX & VXM \\
US treasury rate 2y & TU\\
US treasuret real rate 10y & TY \\
30d Federal Fund rate & FF \\
Gold & GC \\
Silver & SI \\
Copper & HG \\
Crude & CL\\
\hline
  \end{tabular}
\end{table}

\begin{figure}
	\centering
	\caption{Choice of the fractional parameter by the \protect\citet{geweke2008} parameter. The red dotted line represents the one used by a typical first order integration.}\label{fig:1}
	\includegraphics[width=0.65\linewidth]{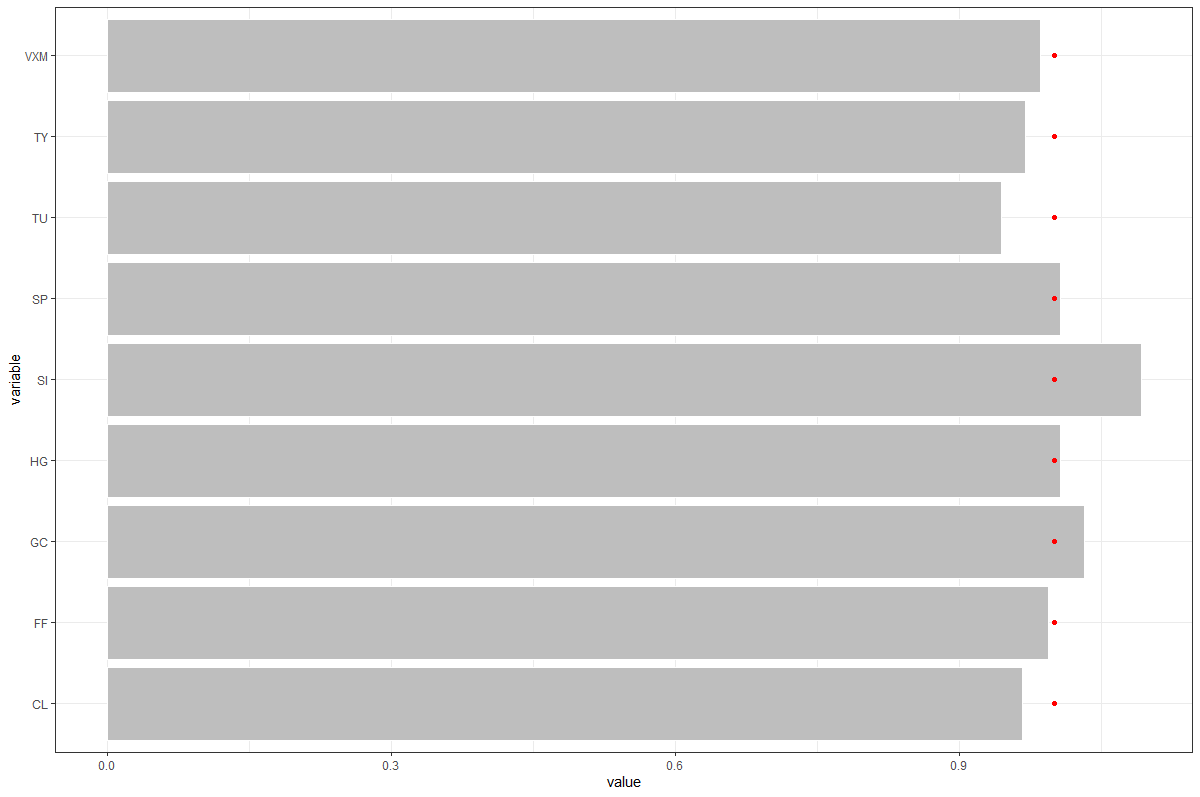} 
\end{figure}

In a first step, time series are fractionally differentiated to achieve stationarity. To estimate the fractional parameter, we use the algorithm introduced by \citet{geweke2008}, whose estimator is based on a regression equation that uses the periodogram function as an estimate of the spectral density. Figure~\ref{fig:1} shows the fractional parameter chosen by the \citet{geweke2008} parameter while the red dotted line represents the one used by a typical first order integration. As it is visible, by using a simple first order integration some variables would be over-integrated and some under-integrated.

We apply the three methodologies on monthly realized covariances calculated as the cross-products of intradaily returns successively aggregated at monthly frequency, following \citet{Barndorff-Nielsen2004}. This procedure ensures that the realized covariance matrix are at least positive definite.

In order to keep the model as parsimonious as possible, the dependent variables of the VLSTAR and TVAR models are the first three principal components of the realized covariances. Instead, the hierarchical clustering model must be applied on metrics since they induce an intuitive topology on a given set which allows to define the notion of distance and similarity. Since covariance is not a metric,  the covariance matrices are converted into correlations and then the entries are adjusted as
\begin{equation*}
d_{i,j}=\ 1-{{\rho }_{i,j}}^2
\end{equation*}
to obtain a distance that satisfies the three axioms of a Euclidean metric \citep[see][]{Osterwalder1975}. Finally, the principal components of the metrics are used as the basis for the AGNES unsupervised learning algorithm.

\begin{figure}
  \centering
  \caption{Clusters sorted by their dissimilarity values. }\label{fig:2}
 \includegraphics[width=0.6\linewidth]{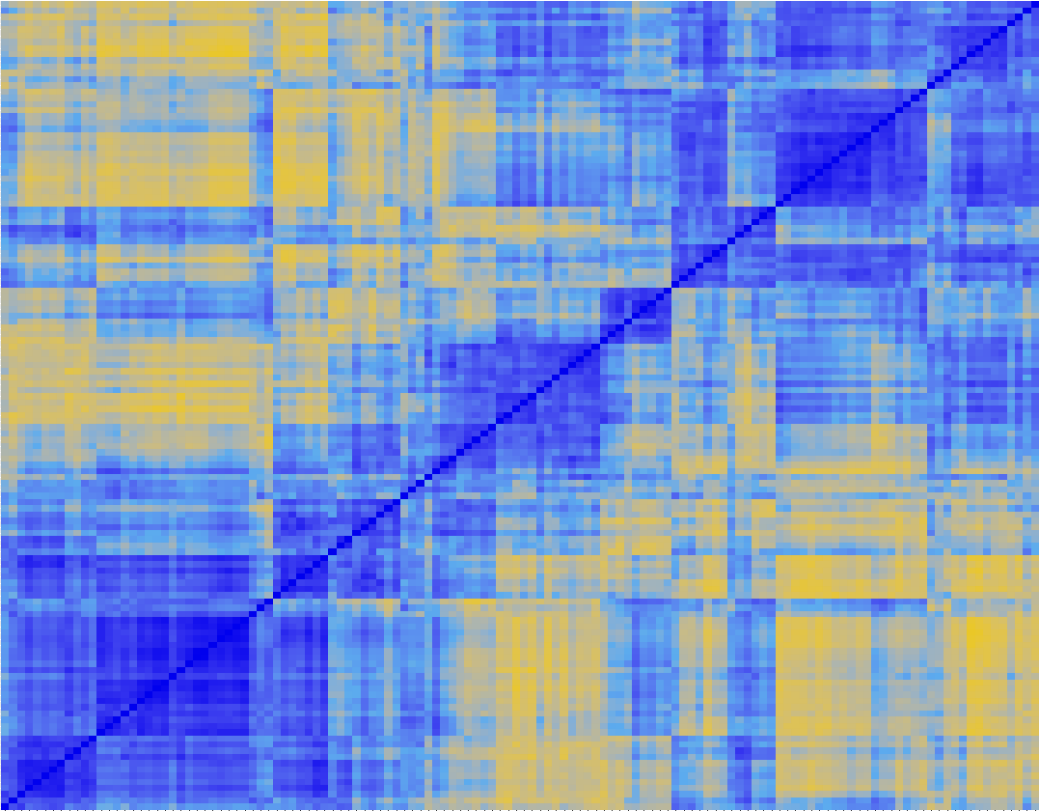} 
\end{figure}

Before clustering, we normalize the time series of metrics extracted from the covariance matrices since clustering techniques are scale sensitive due to the reliance on distance functions. Afterwards, we assess the clustering tendency of the data. To do so, we first calculate the Hopkins’ statistic on the spatial randomness of data, which compares the distances between a sample of data points and their nearest neighbours to the distances from a sample of pseudo points and their nearest neighbours. Under the null hypothesis that the data is un-clusterable, the test statistic follows a beta distribution with both parameters equal to the number of points selected to sample the observations. The closer the Hopkins’ statistics is to 1 and far above 0.5, the more the object have higher tendency to cluster. Our test shows a value of 0.94, hence rejecting the null hypothesis of un-clusterable data. Moreover, we couple this statistical approach to a visual one, by plotting the clusters sorted according to their dissimilarity values as visible in Figure~\ref{fig:2}. As such, if there is clustering tendency, the plot shows several square-shaped dark blocks along the main diagonal. As it is visible from the Figure~\ref{fig:2}, two to three clusters are clearly identifiable along the main diagonal. This further supports our choice of working with two market regimes.

\begin{figure}
  \centering
  \caption{Dendogram produced by the AGNES algorithm}\label{fig:3}
  \includegraphics[width=0.65\linewidth]{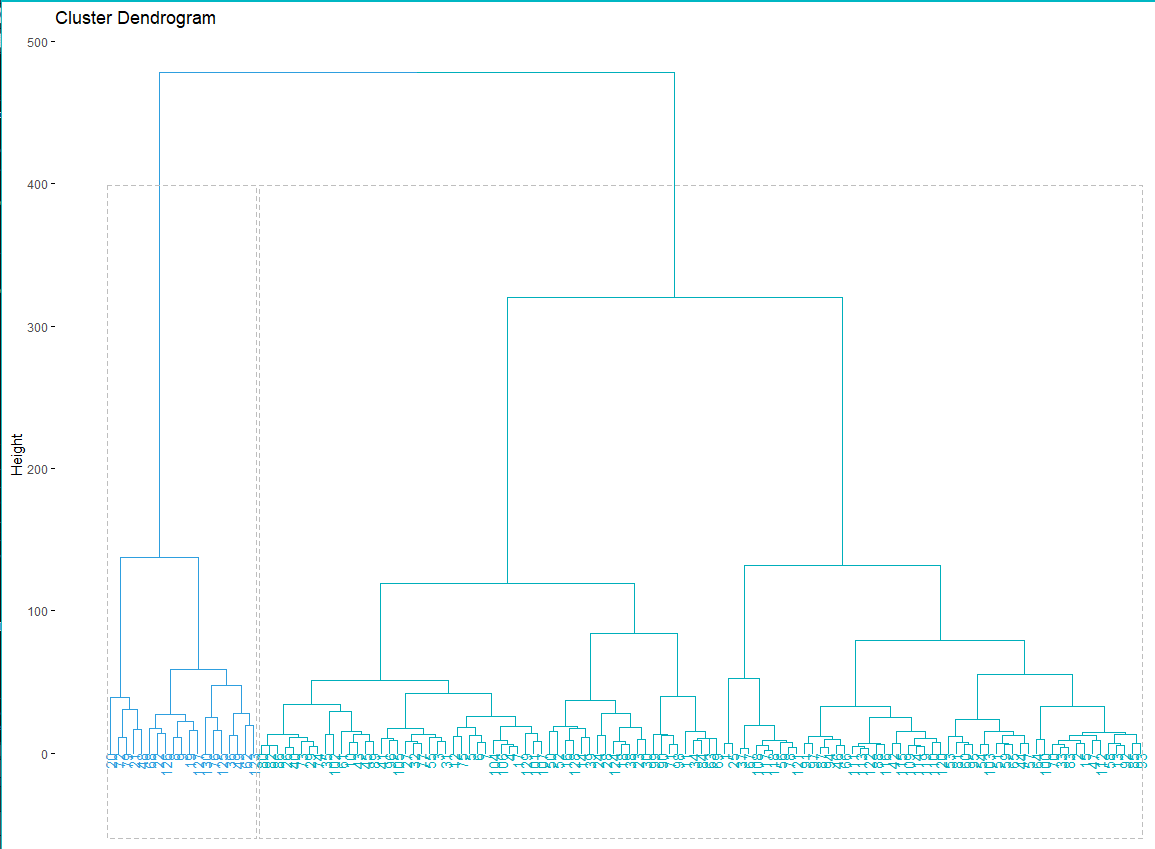} 
\end{figure}

Having assessed the tendency to cluster, we run the AGNES algorithm on the first three principal components of the metrics, since they explain almost 81\% of the variance. As specified above, we use a Manhattan distance and the Ward linkage function. The resulting dendogram is showed in the figure~\ref{fig:3}. 

\begin{figure}
  \centering
  \caption{Clusters in terms first and second principal components.}\label{fig:4}
 \includegraphics[width=0.65\linewidth]{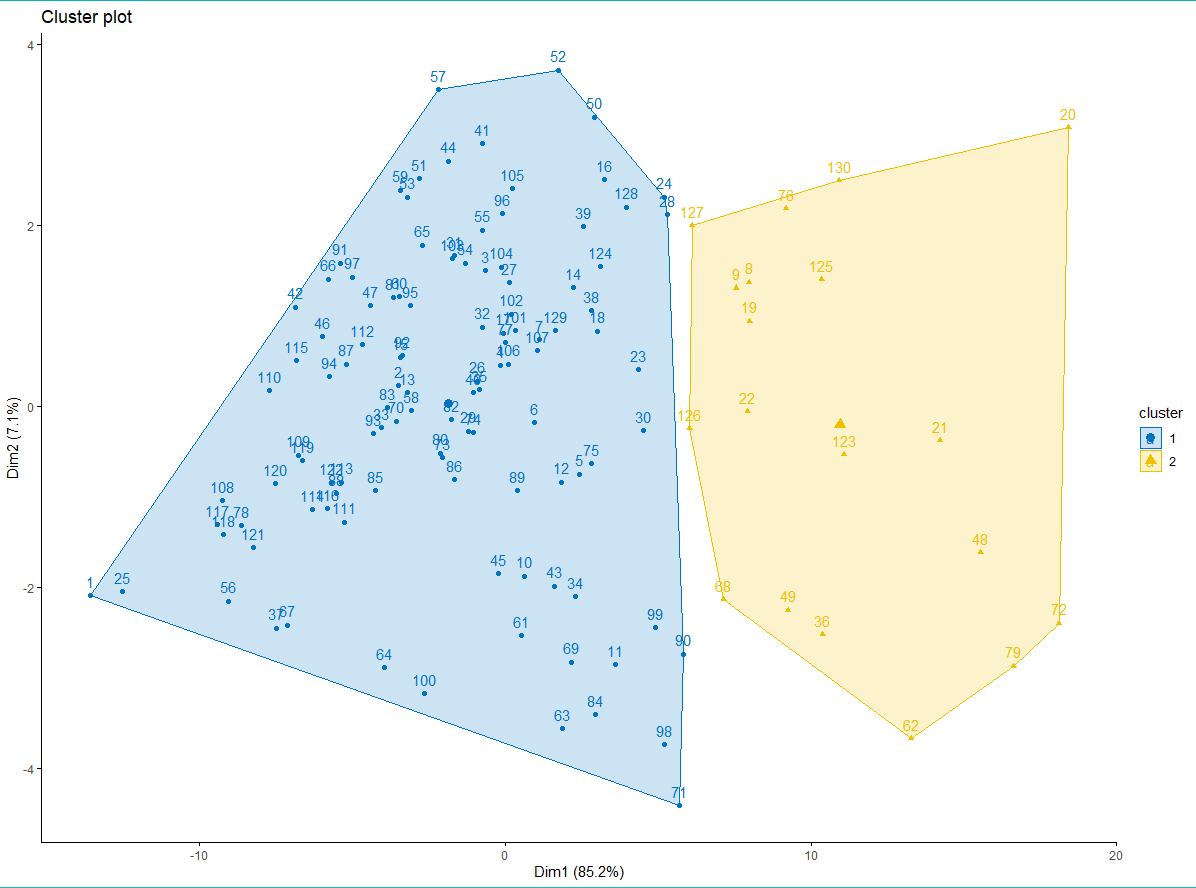} 
\end{figure}

Cutting the dendogram at two clusters yields a clear division between the two market regimes as shown in Figure~\ref{fig:4}. Each dot represents an observation of the first two principal components with those in the yellow area classified into the highly volatile regime while those in the blue area in the calm one. As it is visible and empirically found in financial markets (see also \citet{Fleming2011}), the clustering yields a class imbalance with a lower number of months in the highly volatile regime.

Finally, we validate the clustering by using the silhouette plot, which is a measure of how similar an object is to its own cluster (cohesion) compared to other clusters (separation). The silhouette ranges from $-1$ to $+1$, where a high value indicates that the object is well matched to its own cluster and poorly matched to neighbouring clusters. If most objects have a high value, then the clustering configuration is appropriate. If many points have a low or negative value, then the clustering configuration may have too many or too few clusters. The result of this test highlights that only 3.6\% of the clusters have negative silhouette values hinting that their classification is less appropriate. Further inspection reveals that these are the objects closest to the transition area between the two regimes. Moreover, as a further statistical measure to validate the clusters we use the Dunn index, calculated as the ratio between the minimum inter-cluster distance and the maximal intra-cluster distance. A lower value implies that the clusters are neatly separated. Our procedure obtains a Dunn index of 0.1, indicating a good clusterization.

The VLSTAR model is applied to the principal components of the $\left(T \times N\cdot(N+1)/2\right)$ matrix of realized covariances. We use the first three principal components as they capture almost 85\% of the variance. In order to choose the transition variable, we test the linearity of the model for each candidate transition variable and select the one exhibiting the lowest $p$-value. In this framework, we use a linear trend and the first lag of the principal components as candidates, and we observe the lowest p-value ({\textless}1\%) is associated with the first principal component as transition variable. Thus, the lag of the first principal component is used to drive the dynamics of the VLSTAR model.

\section{Validation}

\begin{figure}
  \centering
  \caption{S\&P500 monthly returns and (in grey-shaded areas) regimes detected by the three models.}\label{fig:5}
 \includegraphics[width=\linewidth]{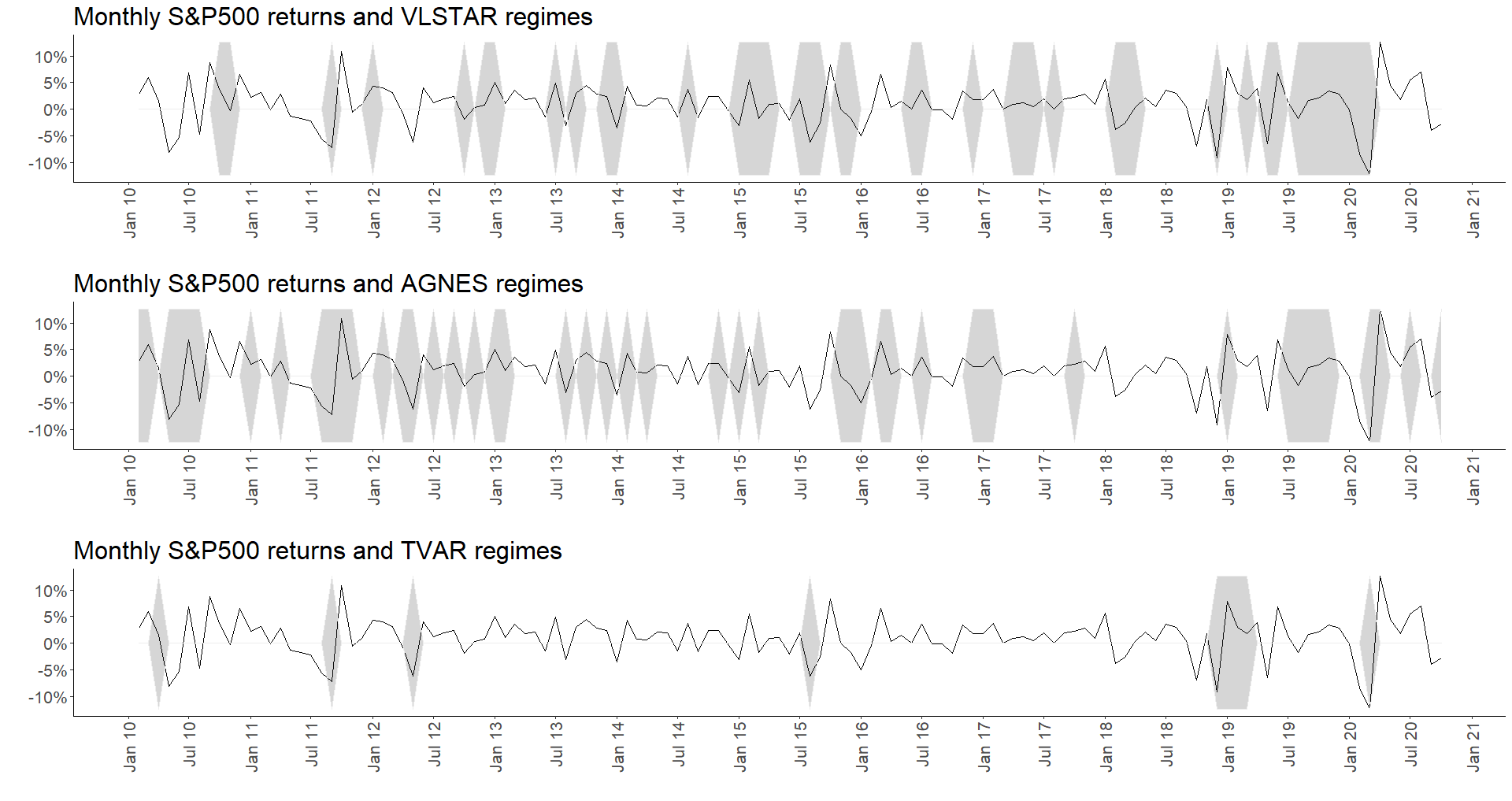} 
\end{figure}

Since market regimes are latent unobservable processes, the detected regimes should be validated in an out-of-sample framework. As a preliminary analysis, the regimes detection by the three methods is visually validated in Figure~\ref{fig:5}, which shows the time series of the monthly S\&P500 returns with the grey shaded areas representing a highly volatile regime detected by each methodology. All the methodologies detect a regime change at the onset of COVID-19 outbreak occurred in March 2020. During this period, in fact, financial markets abruptly contracted amid fears of liquidity crashes and of consumption and production cycles swiftly distorted due to the confinement measures. Ever since the second half of 2019, financial investors were cautiously monitoring the development of geo-political conflicts such as the USA-China trade war and Brexit. Both the VLSTAR and AGNES detect regime changes during those periods. The TVAR, on the other hand, detects only few regime changes, possibly due to its static threshold. The VLSTAR also detects regime changes in line with most of the large S\&P500 drops, such as those in August 2011, July 2015 and January 2018. In particular, the VLSTAR signals a highly volatile regime almost for the whole 2015-2016 biennium, which includes the global decline in stock prices resulted by the sensible slowdown in Chinese GDP growth leading to the SSE Composite Index to fall by 43\% in just over two months and to the deep-seated devaluation of the yuan. Moreover, this two-year period is also characterized by the Brexit referendum in July 2016. The Brexit referendum is caught by the VLSTAR and the AGNES, while the TVAR does not spot a regime change. The AGNES algorithm seems to overreact during periods of market lateral movements, especially during the period July 2011 – December 2012. During this period, the VLSTAR reacts to the market sell-off in August 2011, during which the S\&P500 contracted by 6.7\% in one month with all the 500 stocks falling. Finally, the VLSTAR is the only methodology that reacts to the market decline that began on January 26th 2018, triggered by comments from the US Treasury Secretary Mnuchin suggesting that the government would let the value of the US Treasuries fall as the country would be seeking additional funding for its expansionary plans regarding infrastructure rebuilding and defence. 

To analytically validate the regime detection methodologies, we propose a two-pillars validation framework.
As first valuation approach, we apply the three methodologies to a randomly generated multivariate time-series dataset with controlled variance. In fact, by keeping the mean stable and by modifying the variance of a random data generation process, we are capable to identify the timestamp belonging to a highly volatile regime. The algorithm used for generating the returns of $n = 5$ assets is the following.
 
\begin{enumerate}
\item A vector of $T \time 1$ vector realizations is simulated, and saved in the object $b$, from an AR(1) model with an autoregressive coefficient, $\alpha$, equal to $0.9$, and standard deviation of the white noise equal to $0.1$.
\item Simulated $b$ observations are normalized through:

\begin{equation*}
\overline{b} = \frac{b - min(b)}{max(b)-min(b)}
\end{equation*}

to obtain a vector of time dependent data with values between 0 and 1.
\item The index $t^*$ of the time series $\overline{b}$ is saved such that:

\begin{equation*}
	t^* =
	\begin{cases}
		1 & \text{if $\overline{b}_t > 0.7$}\\
		0 & \text{if $\overline{b}_t \leq 0.7$}
	\end{cases}       
\end{equation*}

\item We simulate $T \times n$ observations from a multivariate Normal distribution with mean vector $0$ and variance matrix $\Sigma$ with diagonal values equal to $0.3$ and off-diagonal values equal to $0.1$ and store the simulated data in the object $Z = \{z_1', \ldots, z_T'\}$.
\item The squares of the first row of $Z$ are saved in $\hat{\sigma}^2_{j,1}$.
\item An error is generated as:

\begin{equation*}
e_{j,t} =
	\begin{cases}
		3\cdots \hat{\sigma}^2_{j, t-1} Z_{j,t} & \text{if $\overline{b}_t > 0.7$}\\
		\hat{\sigma}^2_{j, t-1} Z_{j,t} & \text{if $\overline{b}_t \leq 0.7$}
	\end{cases}       
\end{equation*}

where $t = 1, \ldots, T$ and $j = 1, \ldots, 5$.
\item Returns are generated as:

\begin{equation*}
r_{j,t} = \mu + \phi r_{j, t-1} + e_{j,t}
\end{equation*}

where $\phi = 0.2$.
\item Update values of $\hat{\sigma}^2$ as:

\begin{equation*}
\hat{\sigma}_{j,t}^2 = \omega + \beta \hat{\sigma}_{j,t-1}^2.
\end{equation*}

where $\omega= 0.3$ is the mean variance and $\beta = 0.55$ is the Markov dependence coefficient on the variance in the previous period.

\end{enumerate}

This algorithm permits to cluster volatility in pre-defined dates and to know which of the two regimes of volatility is observed in every observation. To scrutinize the performance on the randomly generated data, a classification-like process based on a confusion matrix is implemented. As such, we first calculate the accuracy of each methodology by adding the correct classifications of the calm and highly volatile regimes. Moreover, it is well-known that classifying a highly volatile regime as a calm one is a more costly misclassification compared to classifying a calm regime as a highly volatile one. In the former case, in fact, an investor would fail to properly hedge its portfolio against tail-risk. Since the loss side of the volatility-return relationship is empirically more convex than the profit side, we deem the likely loss from the incorrect classification of a highly volatile regime as costlier. 

Table~\ref{tab:2} shows the confusion matrices of the three methodologies applied to the randomly generated data. The VLSTAR exhibits the highest accuracy at almost 81\%, while the TVAR the lowest at 35\%. The hierarchical clustering performs in between with an overall accuracy of almost 64\%. The ranking does not capsize when looking at the misclassification of highly volatile regimes. In fact, the VLSTAR misclassifies 4\% of these observations, while the TVAR a high 57\%.

\begin{table}
  \centering
  \caption{Confusion matrices produced by applying the three procedures to the randomly generated data.}\label{tab:2}
  \begin{tabular}{*{12}c}
    \multicolumn{4}{c}{\bfseries VLSTAR(1)} & \hskip1cm& \multicolumn{3}{c}{\bfseries Hierarchical clustering} &\hskip1cm & \multicolumn{3}{c}{\bfseries TVAR (1)}\\\cline{1-4}\cline{6-8}\cline{10-12}
    &&\multicolumn{2}{c}{Predicted} & &  & \multicolumn{2}{c}{Predicted} && & \multicolumn{2}{c}{Predicted}\\
    && Calm & High-Vol &&&Calm&High-Vol&&&Calm&High-Vol\\\cline{3-4}\cline{7-8}\cline{11-12}
    \multirow{2}{*}{Realized}& Calm &\cellcolor{green!30} 65\% & \cellcolor{orange!30} 15\% & &Calm & \cellcolor{green!30} 45\% &\cellcolor{orange!30} 11\% & & Calm & \cellcolor{green!30} 32\% & \cellcolor{orange!30} 8\%\\
    & High-Vol &\cellcolor{red!30} 4\% & \cellcolor{green!30} 16\% & & High-Vol & \cellcolor{red!30} 25\% &\cellcolor{green!30} 19\% && High-Vol & \cellcolor{red!30} 57\% & \cellcolor{green!30} 3\%\\
\cline{1-4}\cline{6-8}\cline{10-12}
  \end{tabular}  
\end{table}

As a second validation tool, we consider a na\"{i}ve momentum investment strategy that buys the underlying when the thirty days moving average is above the one calculated on hundred days. Typically,  momentum investment strategies perform well when financial markets are in calm regime, while they tend to fail when volatility jumps higher.  A regime-based filter is added to this strategy such that there is no trade when a transition towards a highly volatile regime is forecast. As such, we expect the investment strategy not only avoiding long equity positions during market downturns, but also to decrease the downside volatility and to reduce the exposure to liquidity crises.  

\begin{figure}
  \centering
  \caption{Equity line for each investment strategy.}\label{fig:7}
 \includegraphics[width=0.9\linewidth]{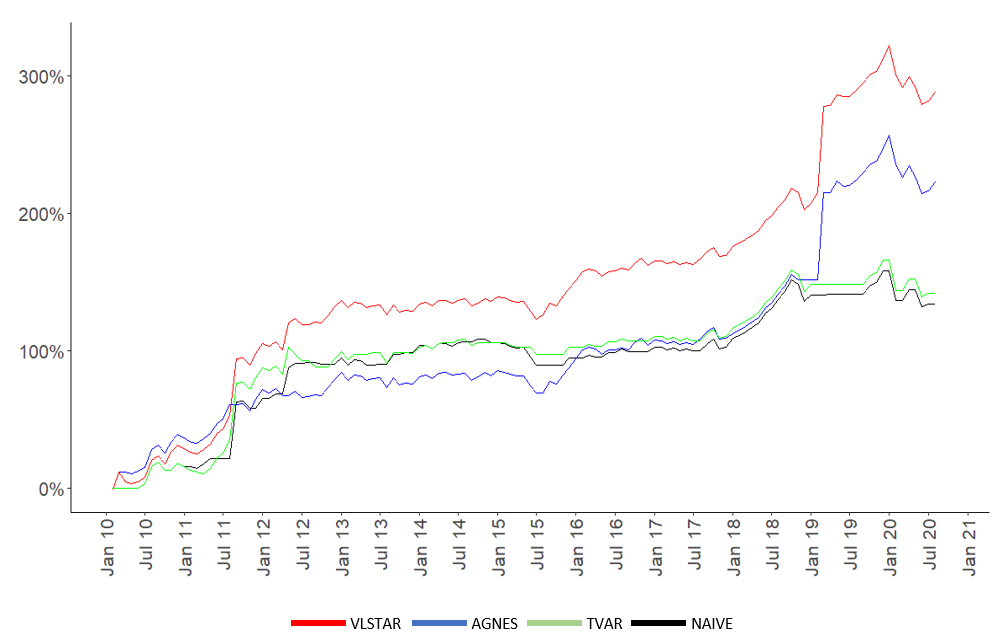} 
\end{figure}

Figure~\ref{fig:7} shows the cumulated equity line of the na\"{i}ve and the filtered investment strategies. Both the VLSTAR and the AGNES filtered investment strategies improve the performance of the investment strategy. In particular, it is visible how these filters allow to avoid the deepest point of the market sell-off and to correctly time the new long entry in the equity markets. In particular, strong accelerations are visible in January 2011 and January 2018. The equity line of the TVAR strategy is very close to the na\"{i}ve’s one since it detects only few market regime changes.

Moreover, the VLSTAR and AGNES out-perform the na\"{i}ve strategy also in terms of annualized Sharpe ratios as reported in the table ~\ref{tab:6} below.

\begin{table}
  \centering
  \caption{Annualized Sharpe ratios of each filtered strategy compared to the na\"{i}ve one.}\label{tab:6}
  \begin{tabular}{lr}
    &\textbf{Sharpe Ratio}\\
    \hline
    Na\"{i}ve & 0.65\%\\
    VLSTAR & 0.93\%\\
    Hierarchical clustering & 0.82\%\\
    TVAR & 0.65\%\\
    \hline
  \end{tabular}
\end{table}

%

Finally, we estimate the transaction costs as a function of the intradaily volatility as follows: 

\begin{equation*}
\mathit{TC}_t=\lambda _0+\lambda _1^{-1}\sigma _t
\end{equation*}

where  $\mathit{TC}_t$  represents the transaction cost at time $t$,  $\sigma_t$  the intradaily volatility,  $\lambda_0$  the fixed portion of the trading costs and  $\lambda_1^{-1}$  the variable one dependent on the liquidity level in order to resemble the effect of a larger downside channel which occurs in such cases. Table \ref{tab:7} below reports the annualized transaction costs for each strategy. The filtered strategies record lower costs not only because of the lower number of trades due the no-trading period in highly volatile regimes, but also because it avoids trades when the intraday volatility is higher as well as the markets are less liquid. We use a value of 1 basis point for  $\lambda_0$  and 0.5 basis point for  $\lambda_1^{-1}$. 

\begin{table}
  \centering
  \caption{Total transaction costs (in basis points) incurred by each filtered strategy compared to the na\"{i}ve one.}\label{tab:7}
  \begin{tabular}{lr}

    &\textbf{Transaction costs in basis points}\\
      	\hline
    Na\"{i}ve & 139.1\%\\
    VLSTAR & 122.5\%\\
    Hierarchical clustering & 132.6\%\\
    TVAR & 138.7\%\\
    \hline
  \end{tabular}
\end{table}

\section{Conclusion and Future Work}

The aim of this paper is to label two market regimes – a calm and a highly volatile – at monthly frequency starting from realized covariance matrices. In particular, the work here presented focuses on the application and the comparison of regime-detection performance of a vector logistic smooth transition autoregressive model (VLSTAR) and the agglomerative hierarchical clustering applied on metrics derived from the covariances. Both models are suitable to regime detection as while the hierarchical cluster bases the regime switches by identifying patterns based on the distance between variables, while the VLSTAR uses a cumulative logistic function to assess the effect of a transition variable on a selected number of regime changes. Moreover, the results from the two main methods are compared with those of a Threshold vector autoregressive model (TVAR). 

Since market regimes’ switches are a latent unobservable process, to assess the validity of our approach we implement a validation framework based on two pillars. First, we apply the regime detection methodologies to a randomly generated time series with controlled variance. Second, we compare the performance of a na\"{i}ve investment strategy to a regime-filtered one. 

The results highlight that the VLSTAR is the best performing model for labelling market regimes, as it achieves the highest accuracy when applied to the randomly generated data as well as it minimizes the misclassification of highly volatile regimes as calm ones. Moreover, it is capable to improve a na\"{i}ve investment strategy with respect to Sharpe ratio and transaction costs. The TVAR model, instead, performs the worst as it does not react to regime switches. The performance of the hierarchical clustering is acceptable, yet outperformed by the VLSTAR. 

Future works should focus on applying regime labels to traditional topics such as regime dependent asset allocation, regime dependent Value at Risk estimation or regime dependent monetary policy.

\sectionbiblio

\bibliographystyle{agsm}
\newpage
\bibliography{Regime}

\end{document}